\documentclass[dvips]{article}
\usepackage{icrctc07}

\title{Successful ToO triggers on the extragalactic sources with the MAGIC telescope}
\shorttitle{ToO triggers with MAGIC}
\authors{D. Mazin$^{1}$, and E. Lindfors$^{2}$ for the MAGIC collaboration$^{3}$.}
\shortauthors{D. Mazin et al.}
\afiliations{$^1$Max-Planck-Institute for Physics, D-80806 Munich, Germany\\
$^2$Tuorla Observatory, Turku University, FI-21500 Piikki\"o, Finland\\
$^3$collaboration list at: \texttt{http://wwwmagic.mppmu.mpg.de/collaboration/members}}
\email{mazin@ifae.es, elilin@utu.fi}

\abstract{
The MAGIC collaboration has been performing Target of Opportunity (ToO)
observations whenever alerted that known or potential very high energy
gamma-ray emitting extragalactic sources were in a high flux state in the
optical, X-ray band or/and in the TeV energy range. Here we report on MAGIC
observations performed after such triggers, results of the analysis, and a
possible optical-TeV correlation seen in the data. Detections as well as
spectral and temporal characterestics of Mkn 180, PKS\,2155-304, and
1ES\,1011+496 are reported. }

\begin{document}
\maketitle

\section{Introduction}
The search for very high energy (VHE, defined as $E > 100$~GeV)
$\gamma$-ray emission from Active Galactic Nuclei (AGN) is one of the major
goals for ground-based $\gamma$-ray astronomy. New detections open the
possibility of a phenomenological study of the physics inside the
relativistic jets in AGN, in particular to understand both the origin of the
VHE $\gamma$-rays as well as the correlations between photons of different
energy ranges (from radio to VHE). The number of reported VHE $\gamma$-ray
emitting AGN has been slowly increasing and is currently 17 (May 2007).
Eight of them have been seen by MAGIC, all of the them belonging 
to the BL~Lac class of AGNs.

The spectral energy distribution (SED) of BL~Lac objects normally shows a
two bump structure. 
The first peak in high frequency peaked BL~Lacs (HBLs) 
has a maximum in the X-ray band,
whereas the second peak is located in the GeV-TeV
band.  It is believed that the radiation is produced in a highly beamed plasma jet, which is
almost aligned with the observer's line of sight. A double peaked SED is
normally attributed to a population of relativistic electrons, where one
peak is due to synchrotron emission in the magnetic field of the jet, and
the second peak is caused by inverse Compton (IC) scattering of low energy
photons by the same parent relativistic electron population.

The known VHE $\gamma$-ray emitting BL~Lacs are variable in flux in all
wavebands.  Correlations between X-ray-$\gamma$-ray emission have been found
(e.g.  \cite{fossati04}) although the relationship has proven to be rather
complicated with $\gamma$-ray flares being also detected in the absence of
X-ray flares 
\cite{krawczynski421_1959} 
and vice versa
\cite{rebillot421}. The optical-TeV correlation has yet to be studied, but
the optical-GeV correlations seen in 3C~279 \cite{hartman3C279} suggest that
at least in some sources such correlations exist. Using this as a guideline,
the MAGIC collaboration has been performing Target of Opportunity (ToO)
observations whenever being alerted that sources were in high flux state in
the optical and/or X-ray band.

In this paper we present the first detection of VHE $\gamma$-ray
emission from Markarian~180 (Mkn~180), 1ES\,1011+496, and results on PKS 2155-304.

\section{Markarian 180}
\begin{figure}
\begin{center}
\includegraphics [width=0.48\textwidth]{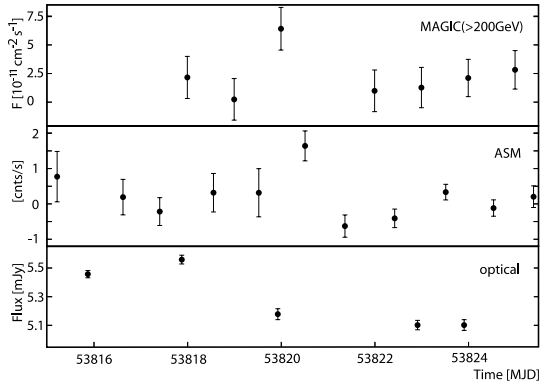}
\end{center}
\caption{\label{fig_LC} Lightcurve of Mkn~180 for 
    MJD=53815-53825 (March 21 to March 31).
    Upper panel: VHE $\gamma$-rays measured by MAGIC above 200~GeV.
    Middle panel: ASM count rate.
    Lower panel: optical flux measured by KVA.} 
\end{figure}

The AGN Mkn~180 (1ES 1133+704) is a well-known high frequency peaked BL Lac
(HBL) at a redshift of $z=0.045$ \cite{falco}. 
The observation of Mkn~180 was triggered by a
brightening of the source in the optical on 2006 March 23.
The alert was
given as the core flux increased by 50\% from its quiescent level value as
determined from over three years of data recording.
The simultaneous MAGIC, ASM\footnote{All-Sky-Monitor on-board the RXTE satellite, 
see http://heasarc.gsfc.nasa.gov/xte\_weather/} 
and KVA\footnote{See http://users.utu.fi/kani/1m/index.html}
lightcurve is shown in Fig.~\ref{fig_LC}.  Around this time Mkn 180 was also
observed as part of the AGN monitoring program by the University of Michigan
Radio Observatory (UMRAO). 
No evidence for flaring was found in the radio data between January and April 2006.

Mkn~180 was observed by MAGIC in 2006 during 8 nights (from March 23 to 31), for a
total of 12.4~hours,  at zenith angles ranging from $39^{\circ}$ to
$44^{\circ}$.  After the standard analysis, a clear excess corresponding to
a 5.5~$\sigma$ excess was determined.  No evidence for flux variability was
found.  The fit to the nightly integrated flux is consistent with a constant
emission ($\chi^{2}=7.1$, 6 degrees of freedom).
Fig.~\ref{fig_LC} shows the VHE lightcurve together with the ASM daily
averages and the R-band by KVA flux data. The X-ray flux of the source is generally
below the ASM sensitivity, but on March 25 a 3~$\sigma$ excess was observed,
which suggests that the source was also active in X-rays.  The optical flux
reached its maximum in the night MAGIC started the observations (March 23)
and began to decrease afterwards.
The measured energy spectrum of Mkn~180 can be well 
described by a power law:
\begin{eqnarray} 
\nonumber
{\frac{\mathrm{d}N}{\mathrm{d}E}}= (4.5\pm1.8)\times10^{-11}\times \\ 
\nonumber
\left({\frac{E}{0.3\,\mathrm{TeV}}}\right)^{-3.3\pm0.7}\,
\frac 1 {\mbox{TeV}\,\mbox{cm}^{2}\,\mbox{s}}  
\end{eqnarray}
The observed integral flux
above 200~GeV is $F(E>200\:\mathrm{GeV})=(2.25\pm0.69)\times10^{-11}
\mbox{cm}^{-2}\mbox{s}^{-1}$, which corresponds to $1.27\times10^{-11}\,
\mathrm{ergs}\, \mathrm{cm}^{-2}\mathrm{s}^{-1} $ 
resp. 11\% of the Crab Nebula flux measured by MAGIC. 
Details on the analysis and the results can be found in \cite{magic180}.

\begin{figure}[t]
\begin{center}
\begin{minipage}[b]{0.45\textwidth}
\includegraphics*[width=0.9\textwidth,angle=0,clip]{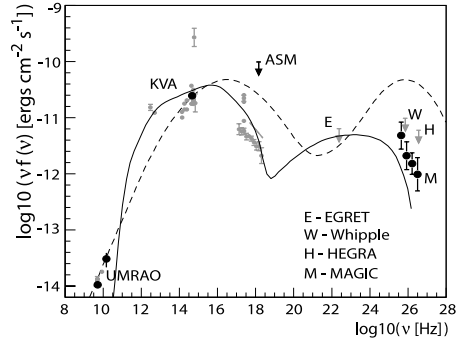}
\caption{\label{fig_SED} The Spectral Energy Distribution of
  Mkn~180. Simultaneous data (UMRAO, KVA, ASM, and MAGIC) are noted 
  in black. Grey points represent historical data. 
  The arrows denote the upper limits from ASM, EGRET, Whipple,
  and HEGRA.  The solid line is from \cite{costamante} and the dashed 
  line is from \cite{fossati}.}
\end{minipage}
\end{center}
\end{figure}

\section{PKS 2155-304}
\begin{figure}
\begin{center}
\includegraphics [width=0.48\textwidth]{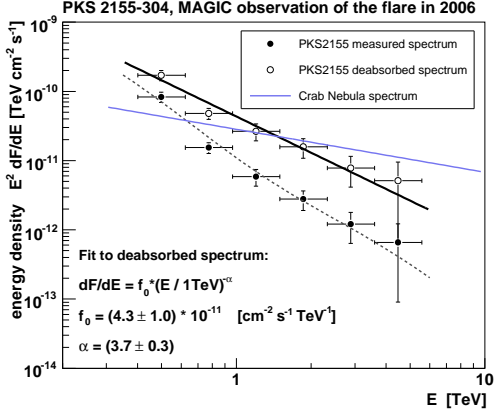}
\end{center}
\caption{\label{fig_pks2155} Averaged energy spectrum of PKS2155-304 for 
    MJD=53944-53949 (2006 July 28 to August 2). The Crab Nebula spectrum is shown for comparison.}
\end{figure}

PKS 2155-304 is a well-known HBL object, located in the Southern 
hemisphere at a redshift of $z = 0.117$ \cite{falomo2155}.
MAGIC observations were triggered by the Astronomer's Telegram 867
\cite{benbow} reporting historically high fluxes of 
VHE $\gamma$-rays from PKS 2155-304. The VHE activity was
supported by an optical brightening of the source and an increased
X-ray activity.

MAGIC observed PKS 2155-304 for 6 nights in July -- August 2006 
at high zenith angles ranging from $58^{\circ}$ to $68^{\circ}$. 
A clear signal corresponding
to 11~$\sigma$ excess was determined. Due to the high zenith of observation,
the  energy threshold is higher.
The analysis energy threshold for this data 
set is at about 600 GeV. An averaged energy spectrum of PKS 2155-304 
is shown in Figure~\ref{fig_pks2155}. The measured spectrum is consistent with a power law with 
a soft slope of $\alpha=4.0$, whereas the deabsorbed energy spectrum \cite{kneiske:2002a}
has a slope of $\alpha=3.7 \pm 0.3$. 
The soft slope confirms findings by H.E.S.S.
\cite{aharonian:2007:hess:2155:flare} and points to a IC peak position below 200 GeV. 
No significant flux variations were
found in 30 min time bins along the 6 nights of observations. 

\section{1ES\,1011+496}

1ES~1011+496 is an HBL object for which
we now determined a redshift of $z=0.212\pm0.002$.
Previously, this has been uncertain since it was based on an
assumed association with the cluster Abell 950 \cite{wisniewski:1985a}. 
The MAGIC observation was triggered by an observed high optical state
of 1ES~1011+496 on 2007 March 12th.

\begin{figure}
\includegraphics[width=0.45\textwidth]{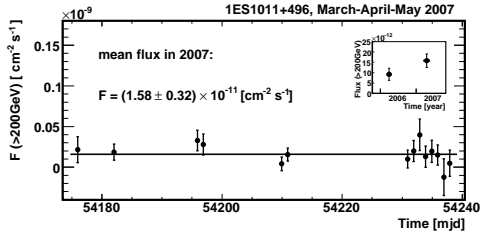}
\label{fig:1011lc}
\caption{The night-by-night light curve of 1ES~1011+496 from 2007 March 17 (MJD 54176) to 2007 May 18 (MJD 54238).
         The year-by-year light curve is shown in the inset, the 2006 data point is from \cite{hbl}.}
\end{figure}

After the alert, MAGIC observed 1ES~1011+496 in March--May 2007.  
The total observation time was 26.2 hours and the observation was
performed at zenith angles ranging from $20^{\circ}$ to
$37^{\circ}$. After removing runs with unusual trigger rates, mostly
caused by bad weather conditions, the effective observational time
amounts to 18.7 hours. Using the standard analysis chain for the MAGIC
telescope data, a signal of 297 events over 1591 normalized background events
was found, corresponding to an excess with significance of $6.2\sigma$.
The averaged energy spectrum of 1ES~1011+496 extends from 
\mbox{$\sim 120$\,GeV} to \mbox{$\sim 750$\,GeV} and can
be well approximated by a power law:
\begin{eqnarray}  
\nonumber
{\frac{\mathrm{d}N}{\mathrm{d}E}}=(2.0\pm0.1)\cdot10^{-10} \times \\
\nonumber
\left({\frac{E}{0.2\,\mathrm{TeV}}}\right)^{-4.0\pm0.5}\,
\frac 1 {\mbox{TeV}\,\mbox{cm}^{2}\,\mbox{s}}  
\end{eqnarray}  
After the correction for the EBL absorption \cite{kneiske:2002a}, 
the slope of the spectrum is $\Gamma_{\mathrm{int}} = 3.3 \pm 0.7$ 
 ($\chi^2\,/\,\mbox{NDF} = 2.55 / 2$), 
softer than observed for other HBLs in TeV energies and thus not providing new
constraints on the EBL density. The energy spectrum can be nicely fitted
by a simple leptonic model \cite{tavecchio:2001a,magic1011}. 
However, we note that due to a lack of simultaneous X-ray and soft $\gamma$-ray
data, the model is degenerated.
To search for time variability the sample was divided into 14
subsamples, one for each observing night. Fig.~\ref{fig:1011lc} shows the integral
flux for each night calculated for a photon flux above 200~GeV. 
The flux is statistically constant at an emission level of
F($>$200~GeV)=(1.58$\pm0.32)\cdot 10^{-11}$ photons cm$^{-2}$s$^{-1}$. 
In the inset of Fig.~\ref{fig:1011lc}, we show integral fluxes
above 200~GeV on a year-by-year basis. The first point in the inset 
corresponds to an earlier observation of 1ES1011+496 by MAGIC in 2006,
which lead to a hint of a signal \cite{hbl}. This hint can 
now be interpreted as being due to a lower flux state of
the source than measured in 2007.  
Details on the analysis and the results can be found in \cite{magic1011}.

\section{Conclusions}

The discovery of VHE $\gamma$-ray emission from Mkn~180 was triggered by an
optical flare, but no significant variations in the VHE regime were found.
The short observation period and the small signal do not allow to carry out
detailed studies. It is therefore not possible to judge whether the detected
VHE flux level represents a flaring or a quiescent state of the AGN.
The rather steep slope of the VHE spectrum of Mkn~180 
suggests an IC peak position well
below 200~GeV, while the non-detection by EGRET gives a lower limit of
$\sim1$~GeV for the peak position \cite{EGRET1}. 
This is in agreement with the leptonic model of
\cite{costamante} suggesting the IC peak position at around 10~GeV.  

The detection of PKS2155-304 with the MAGIC telescope proved the capability
of the instrument to observe distant extragalactic $\gamma$-ray sources 
under large zenith angle if the sources are in a flaring flux state.
Especially for multiwavelength campaigns, MAGIC observations are crucial 
in order to obtain independent energy
spectra and light curves measurements at VHE.

We report the discovery of VHE $\gamma$-ray emission from the HBL
object 1ES~1011+496 in the 2007 data set. With the redshift of $z=0.212$, it is the most
distant source detected to emit VHE $\gamma$-rays up to date.  
The discovery was as well triggered by an optical flare.
The observed spectral properties 
(soft and no significant excess above $\sim 800$~GeV) 
are consistent with the state-of-the art EBL models 
and confirm recently derived EBL limits. 

The discovery of VHE emission from Mkn~180 and 1ES\,1011+496 
during an optical outburst as well as 
an optical activity during the PKS2155-304 VHE outburst make it
very tempting to speculate about the connection between the two energy bands.
Further VHE observations with and without an optical trigger are needed
to prove that there is indeed a correlation. 

\textit{Acknowledgements:}
We would like to thank the IAC for the excellent working conditions at the ORM
in La Palma. The support of the German BMBF and MPG, the Italian INFN, the
Spanish CICYT, the ETH Research Grant TH~34/04~3 and the Polish MNiI Grant
1P03D01028 is gratefully acknowledged.  
\bibliography{icrc0936}
\bibliographystyle{unsrt}

\end{document}